\documentclass[prd,aps,nofootinbib,showpacs,notitlepage,showkeys,preprintnumbers]
{revtex4-1}
\usepackage{graphicx,epsf,amsmath,amsfonts,amssymb,amsbsy}
\usepackage{epsfig}
\usepackage[mathscr]{eucal}
\newcommand{\mat}{\left ( \begin{array}}
\newcommand{\emat}{\end{array} \right )}
\newcommand{\vect}{\left ( \begin{array}{c}}
\newcommand{\evect}{\end{array} \right )}

\newcommand{\qbar}{\overline q}
\newcommand{\Lagr}{\mathscr{L}}

\begin{document}
%\hfill HU-EP-11/11

\title{Finite-size effects in pion condensation
phenomena of dense baryonic matter in the NJL$_2$ model}

\author{D. Ebert$^{1)}$, T.G. Khunjua $^{2)}$, K.G. Klimenko$^{3)}$, and V.Ch. Zhukovsky$^{2)}$}
\affiliation{$^{1)}$ Institute of Physics, Humboldt-University
Berlin, 12489 Berlin, Germany}
\affiliation{$^{2)}$ Faculty of Physics, Moscow State University, 119991, Moscow, Russia}
\affiliation{$^{3)}$ IHEP and University "Dubna" (Protvino branch), 142281, Protvino, Moscow Region, Russia}

\begin{abstract}
The properties of two-flavored massless Nambu-Jona-Lasinio model in
(1+1)-dimensional $R^1\times S^1$ spacetime with compactified space
coordinate are investigated in the presence of isospin and quark
number chemical potentials $\mu_I$, $\mu$. The consideration is
performed in the large $N_c$ limit, where $N_c$ is the number of
colored quarks. It is shown that at $L=\infty$ ($L$ is the length of
the circumference $S^1$) the pion condensation (PC) phase with {\it
zero quark number density} is realized at arbitrary nonzero $\mu_I$
and for rather small values of $\mu$. However, at arbitrary finite
values of $L$ the phase portrait of the model contains the PC phase
with {\it nonzero quark number density} (in the case of periodic
boundary conditions for quark fields). Hence, finite sizes of the
system can serve as a factor promoting the appearance of the PC
phase in quark matter with nonzero baryon densities. In contrast,
the phase with chiral symmetry breaking may exist only at rather
large values of $L$.\end{abstract} \maketitle

\section{Introduction}

Recently, great attention has been paid to the investigation of the
QCD phase diagram in terms of baryonic as well as isotopic (isospin)
chemical potentials. The reason is that dense baryonic matter which
can appear in heavy-ion collision experiments has an evident isospin
asymmetry. Moreover, the dense hadronic/quark  matter inside compact
stars is in general also expected to be isotopically asymmetric. To
describe the above mentioned realistic situations, i.e. when the
baryonic density is comparatively low, usually different
nonperturbative methods or effective theories such as chiral
effective Lagrangians and especially  Nambu -- Jona-Lasinio (NJL)
type models \cite{njl} are employed. In this way the QCD phase
diagram including chiral symmetry restoration
\cite{asakawa,ebert,sadooghi,hiller,boer}, color superconductivity
(CSC) \cite{alford,klim,incera}, and charged pion condensation (PC)
phenomena \cite{son,ek,ak,andersen} were investigated under
heavy-ion experimental and/or compact star conditions, i.e. in the
presence of such external conditions as temperature, chemical
potentials and possible external (chromo)magnetic fields (see the
above references).

Moreover, it was also realized in the framework of NJL-type models
that different finite size factors, i.e. curvature or nontrivial
spacetime topology as well as a finite spatial volume of a system,
can significantly change  the properties  of both the chiral- and
color superconductivity phase transitions. This conclusion is the
result of chiral symmetry breaking investigations in weakly curved
spaces \cite{odin,eliz} and in spaces with nontrivial topology,
where one or more space coordinates are compactified \cite{gusynin}.
In addition, the properties of finite size normal quark matter
droplets in the language of the MIT-bag model were considered, e.g.,
in the review \cite{madsen}. It was also noted that the position of
the chiral critical end point of the QCD phase diagram, which could
be investigated in heavy ion collision experiments, depends
essentially on the size of a finite system  \cite{fraga}. Next, the
effect of spaces with constant curvature or nontrivial topology on
CSC was studied in \cite{etz,ek2}. In particular, there it was shown
that in a restricted region the CSC phenomenon might occur much
easier than in an infinite one \cite{ek2}. Moreover, the stability
of finite size quark matter droplets in the color-flavor locked
phase was considered in the framework of a bag model using the
so-called multiple expansion method \cite{kiriyama}. However, up to
now we have only scarce information about the formation of the pion
condensation phase in dense baryonic matter under the influence of
finite-size effects (see below). This fact is  partially explained
by the complexity of the problem arising from the necessity to
introduce an additional isotopic chemical potential for the adequate
description of quark matter with isospin asymmetry and, in
particular, of the PC phenomenon.

Obviously, the (3+1)-dimensional NJL models depend on the cutoff
parameter which is typically chosen of the order of one GEV, so that
the results of their usage are valid only at {\it comparatively low
energies, temperatures and densities (chemical potentials)}.
Besides, there exists also a class of renormalizable theories, the
(1+1)-dimensional chiral Gross-Neveu (GN) type models \cite{gn},
\footnote{Below we shall use the notation ``NJL$_2$ model''  instead
of ``chiral GN model'' for (1+1)-dimensional models with {\it a
continuous chiral symmetry}, since the chiral structure of the
Lagrangian is the same as that of the (3+1)-dimensional NJL model.}
that can be used as a laboratory for the qualitative simulation of
specific properties of QCD at {\it arbitrary energies}.
Renormalizability, asymptotic freedom, as well as the spontaneous
breaking of chiral symmetry (in vacuum) are the most fundamental
inherent features both for QCD and all GN type models. In addition,
the $\mu-T$ phase diagram is qualitatively the same for the QCD and
GN model \cite{wolff,barducci,chodos,thies} (here $\mu$ is the quark
number chemical potential and $T$ is the temperature). Note also
that the GN type models are suitable for the description of physics
in quasi one-dimensional condensed matter systems like polyacetylene
\cite{caldas}. It is currently well understood (see, e.g., the
discussion in \cite{barducci,chodos,thies}) that the usual no-go
theorem \cite{coleman}, which generally forbids the spontaneous
breaking of any continuous symmetries in two-dimensional spacetime
does not work in the limit  $N_c\to\infty$, where $N_c$ is the
number of colored quarks. This follows from the fact that in the
limit of large $N_c$ the quantum fluctuations, which would otherwise
destroy a long-range order corresponding to a spontaneous symmetry
breaking, are suppressed by $1/N_c$ factors. Thus,  the effects
inherent for real dense quark matter, such as  color
superconductivity (spontaneous breaking of the continuous color
symmetry) or charged pion condensation (spontaneous breaking of the
continuous isospin symmetry) might be simulated in terms of a
simpler (1+1)-dimensional GN-type model, though only in the leading
order of the large $N_c$ approximation (see, e.g.,
\cite{chodos,abreu} and \cite{ekzt,massive}, respectively). Finally,
one should recall that both the chiral phase transition \cite{kim}
and color superconductivity \cite{abreu} were investigated in the
framework of GN models with account of the nontrivial $R^1\times
S^1$ spacetime topology.

In general, this paper is devoted to the consideration of the
charged pion condensation phenomenon under the influence of
finite-size effects. The problem was partially solved in
\cite{universe}, where PC was studied in the framework of the
(3+1)-dimensional NJL model in the Einstein universe with a constant
curvature and finite spatial volume, and in \cite{ekzt}, where it
was considered in the $SU_L(2)\times SU_R(2)$ symmetric GN model in
spacetime with nontrivial topology, i.e. on the $R^1\times S^1$
manifold with compactified space coordinate, and in the assumption
that $N_c\to\infty$. For technical simplifications, in those papers
the quark chemical potential $\mu$ was assumed to be zero, i.e. the
considerations were performed at nonzero isospin $\mu_I$ chemical
potential only (this situation corresponds to quark matter with zero
baryon density). Obviously, the considered problem deserves further,
more detailed investigations, this time when both chemical
potentials are nonvanishing, $\mu\ne 0,\mu_I\ne 0$. This is
physically motivated by the fact that quark matter, which might be
created in heavy-ion collisions, has in general nonzero baryon- and
isospin densities and must be investigated in the framework of a
theory with nonzero $\mu$ and $\mu_I$. Moreover, quark-matter lumps
occupy a finite volume. Hence, in this paper we study, for
illustration, the PC phenomenon in a $SU_L(2)\times SU_R(2)$
symmetric two-dimensional NJL$_2$ model with $\mu\ne 0,\mu_I\ne 0$,
when the spatial coordinate is compactified and $N_c\to\infty$. In
particular, we shall demonstrate that the finite size of the system
promotes the appearance of the PC phenomenon in dense baryonic
matter. We hope that the results of such a study of a renormalizable
two-dimensional model may provide an additional stimulus for further
investigations of the charged PC phenomenon in more realistic
models.

The paper is organized as follows. In Section II we derive, in the
leading order of the large $N_c$-expansion, the expression for the
thermodynamic potential of the two-flavored massless NJL$_2$ model
with quark number chemical potential $\mu$ and isospin chemical
potential $\mu_I$ for zero temperature in $R^1\times R^1$- and in  
$R^1\times S^1$ spacetimes.Then, in
Section III the phase structure of the model is investigated both at
a finite value of $L$ and at $L\to\infty$, where $L$ is the radius
of the $S^1$ circumference. Finally, Section IV presents a summary
and some concluding remarks.

\section{Description of the model and its thermodynamic potential }

\subsection{Effective action}

We consider a (1+1)-dimensional NJL$_2$ model to mimic the phase
structure of real dense quark matter with two quark flavors (u and d
quarks). Its Lagrangian has the form:
\begin{equation}
\begin{split}
\label {2LagrGN1}
\Lagr_{q,\qbar}=\overline{q}\left[\gamma^\nu
  \mathrm{i}\partial_\nu+\mu\gamma^0+\frac{\mu_I}{2}\tau_3\gamma^0\right]q+\frac{G}{N_c}\left[(\overline{q}q)^2+(\overline{q}\mathrm{i}\gamma^5
\vec\tau q)^2
\right],
\end{split}
\end{equation}
where each quark field $q(x)\equiv q_{i\alpha}(x)$ is a flavor
doublet $(i=1,2$ or $i=u,d)$ and color \mbox{$N_c$-plet}
$(\alpha=1,\ldots,N_c)$. Moreover, it is a two-component Dirac
spinor (in (\ref{2LagrGN1}) the summation over flavor, color, and
spinor indices is implied); $\tau_k(k=1,2,3)$ are Pauli matrices.
The quark number chemical potential $\mu$ in (\ref{2LagrGN1}) is
responsible for the nonzero baryonic density of quark matter,
whereas the isospin chemical potential $\mu_I$ is taken into account
in order to study properties of quark matter at nonzero isospin
densities (in this case the densities of $u$ and $d$ quarks are
different). The Dirac gamma matrices in two-dimensional spacetime
have the following form:
\begin{equation}
\begin{split}
\gamma^0=\sigma_1=
\begin{pmatrix}
0&1\\
1&0\\
\end{pmatrix};\qquad
\gamma^1=\mathrm{i}\sigma_2=
\begin{pmatrix}
0&1\\
-1&0\\
\end{pmatrix};\qquad
\gamma^5=\sigma_3=
\begin{pmatrix}
1&0\\
0&{-1}\\
\end{pmatrix}.
\end{split}
\end{equation}

Let us consider the symmetries of the Lagrangian. If $\mu_I=0$, the
Lagrangian (\ref{2LagrGN1}) is not only $SU(N_c)$ symmetric, but
also invariant under transformations of the chiral $SU_{L}(2)\times
SU_{R}(2)$ group. However, if $\mu_I\neq0$, the latter symmetry is
reduced to $U_{I_{3}L}(1)\times U_{I_{3}R}(1)$, where $I_3=\tau_3/2$
is the third component of the isospin operator (here and above the
subscripts $L,R$ mean that the corresponding groups only act on
left- and right-handed spinors, respectively). Obviously, this
symmetry can also be presented as $U_{I_3}(1)\times U_{AI_3}(1)$,
where $U_{I_3}(1)$ is  a vector isospin subgroup, and $U_{AI_3}(1)$
is an axial isospin subgroup. Quarks are transformed under these
subgroups as \mbox{$q\rightarrow\exp(i\alpha\tau_3)q$ and
$q\rightarrow\exp(i\alpha\gamma^5\tau_3)q$}, respectively.

The linearized version of Lagrangian (\ref {2LagrGN1}), which contains
composite bosonic fields $\sigma(x)$ and $\pi_a(x)(a=1,2,3)$, has the
following form:
\begin{equation}
\begin{split}
\label {2LagrGN2}
\Lagr_{\sigma,\pi}=\qbar\left[\gamma^\nu \mathrm{i}\partial_\nu+\mu\gamma^0+\frac{\mu_I}{2}\tau_3\gamma^0-\sigma-\mathrm{i}\gamma^5\pi_a\tau_a\right]q-\frac{N_c}{4G}\left[\sigma\sigma+\pi_a\pi_a\right].
\end{split}
\end{equation}
>From Lagrangian (\ref{2LagrGN2}) one obtains the following constraint
equations for the bosonic fields:
\begin{equation}
\begin{split}
\label {bosons}
\sigma(x)=-2\frac{G}{N_c}(\qbar q);\quad \pi_a(x)=-2\frac{G}{N_c}(\qbar \mathrm{i}\gamma^5 \tau_a q).
\end{split}
\end{equation}
Obviously, Lagrangian (\ref{2LagrGN2}) is equivalent to Lagrangian
(\ref{2LagrGN1}), when using the constraint equations
(\ref{bosons}). Furthermore, it is clear from (\ref{bosons}) that the
bosonic fields are transformed under the isospin $U_{I_{3}}(2)$ and
axial isospin $U_{AI_{3}}(2)$ subgroups in the following manner:
\begin{equation}
\begin{split}
\label {inv}
U_{I_3}(1)&:\quad\sigma\rightarrow\sigma;\quad \pi_3\rightarrow\pi_3;\quad\pi_1\rightarrow\cos(2\alpha)\pi_1+\sin(2\alpha)\pi_2;\quad\pi_2\rightarrow\cos(2\alpha)\pi_2-\sin(2\alpha)\pi_1,\\
U_{AI_3}(1)&:\quad\pi_1\rightarrow\pi_1;\quad \pi_2\rightarrow\pi_2;\quad\sigma\rightarrow\cos(2\alpha)\sigma+\sin(2\alpha)\pi_3;\quad\pi_3\rightarrow\cos(2\alpha)\pi_3-\sin(2\alpha)\sigma.
\end{split}
\end{equation}

To avoid the no-go theorem, which forbids the spontaneous breaking of
continuous symmetries in the considered
case of one space dimension, we restrict the discussion only to the
leading order of the large $N_c$ expansion
(i.e. to the case $N_c\rightarrow\infty$), where this theorem is not
valid \cite{barducci,chodos,thies}. In particular, the effective
action $S_{\mathrm{eff}}[\sigma,\pi_a]$ can be found in this
approximation through the relation:
\begin{equation}
\begin{split}
\label {2Seff1}
e^{\mathrm{i}S_{\mathrm{eff}}[\sigma,\pi_a]}=\int [d\qbar][dq] e^{\mathrm{i}\int d^2x \Lagr_{\sigma,\pi}} .\end{split}
\end{equation}
Carrying out the Gaussian-type path integration in (\ref{2Seff1})
over fermion fields, we obtain
\begin{equation}
\begin{split}
\label {2Seff2}
e^{\mathrm{i}S_{\mathrm{eff}}[\sigma,\pi_a]}=e^{-\mathrm{i}\frac{N_c}{4G}\int d^2x (\sigma^2+\pi_a^2)} \det \mathrm{D}.
\end{split}
\end{equation}
In (\ref{2Seff2}) we have used the notation $\mathrm{D}\equiv
D\times I_c$, where $I_c$ is the unit operator in the
$N_c$-dimensional color space and
\begin{equation}
\begin{split}
\label{D}
D=\mathrm{i}\gamma^\nu\partial_\nu+\mu\gamma^0+\frac{\mu_I}{2}\tau_3\gamma^0-\sigma-\mathrm{i}\gamma^5\pi_a\tau_a.
\end{split}
\end{equation}
Then using the general formula $\det
\mathrm{D}=\exp\mathrm{Tr}_{csfx} \mathrm{ln}\mathrm{D}$ and taking
the logarithm of (\ref{2Seff2}), one obtains the following
expression for the effective action in the leading order of the
$1/N_c$-expansion:
\begin{equation}
\begin{split}
\label{2Seff3}
S_\mathrm{eff}(\sigma,\pi_a)=-N_c\int\frac{\sigma^2+\pi_a^2}{4G}d^2x-\mathrm{i}N_c\mathrm{Tr}_{sfx}\mathrm{ln}D,
\end{split}
\end{equation}
where the $\mathrm{Tr}$-operator stands for the trace in spinor (s),
flavor (f), and two-dimensional coordinate (x) spaces, respectively
(the trace in color (c) space, obviously, equals $N_c$).

\subsection{Thermodynamic potential}

Starting from (\ref{2Seff3}), one can define the thermodynamic
potential (TDP) of the model at zero temperature $T=0$ in the
mean-field approximation, i.e. in the leading order of the
$1/N_c$-expansion \cite{ekzt}:
\begin{equation}
\begin{split}
\label{2Omega1}
\Omega_{\mu\mu_I}(\sigma,\pi_a)\equiv
&\left.-\frac{S_\mathrm{eff}(\sigma,\pi_a)}{N_c\int d^2x}\right|_{\sigma,\pi_a=\mathrm{const}}=\frac{\sigma^2+\pi^2_a}{4G}+\mathrm{i}\frac{\mathrm{Tr}_{sfx}\mathrm{ln}D}{\int d^2x}
\\
=&
\frac{\sigma^2+\pi_a^2}{4G}+\mathrm{i}\mathrm{Tr}_{sf}\int\frac{d^2p}{(2\pi)^2}\mathrm{ln}
(\gamma p+\mu\gamma^0+\frac{\mu_I}2\tau_3\gamma^0-\sigma-\mathrm{i}\gamma^5\pi_a\tau_a),
\end{split}
\end{equation}
where the mean fields $\sigma$ and $\pi_a$ are $x-$independent
quantities, and in the round brackets of (\ref{2Omega1}) just the
momentum space representation $\overline{D}=\gamma
p+\mu\gamma^0+\frac{\mu_I}2\tau_3\gamma^0-\sigma-\mathrm{i}\gamma^5\pi_a\tau_a$
of the Dirac operator $D$ appears. \footnote{There exists also the
interesting case of $x$-dependent mean fields describing chiral
density waves \cite{gubina}. For the following
investigation of finite-size effects we restrict us, however, for
simplicity, to the case of constant fields.} Obviously,
$\mathrm{Tr}_{sf}\mathrm{ln}\overline{D}=\sum_{i}\mathrm{ln}\epsilon_i$,
where summation over all four eigenvalues $\epsilon_i$ of the
$4\times4$ matrix $\overline{D}$ is implied and
\begin{equation}
\begin{split}
\label{epsilon}
\epsilon_{1,2,3,4}=-\sigma\pm\sqrt{(p_0+\mu)^2-p_1^2-\pi_a^2+\left(\frac{\mu_I}2\right)^2\pm\mu_I\sqrt{(p_0+\mu)^2-\pi_1^2-\pi^2_2}}.
\end{split}
\end{equation}
Hence,
\begin{equation}
\begin{split}
\label{2Omega2}
\Omega_{\mu\mu_I}(\sigma,\pi_a)=&\frac{\sigma^2+\pi_a^2}{4G}+i\int\frac{d^2p}{(2\pi)^2}\mathrm{ln}(\epsilon_1\epsilon_2\epsilon_3\epsilon_4)\\
=&\frac{\sigma^2+\pi_a^2}{4G}+i\int\frac{d^2p}{(2\pi)^2}\mathrm{ln}\{[(p_0+\mu)^2-\varepsilon_+^2]\times[(p_0+\mu)^2-\varepsilon_-^2]\},\\
\end{split}
\end{equation}
where
\begin{equation}
\begin{split}
\label{varepsilon}
\varepsilon_\pm=\sqrt{\left(\sqrt{p_1^2+\sigma^2+\pi^2_3}\pm\frac{\mu_I}2\right)^2+\pi^2_1+\pi^2_2}.
\end{split}
\end{equation}
It is clear that the TDP $\Omega_{\mu\mu_I}(\sigma,\pi_a)$ is
symmetric under the transformations $\mu\rightarrow-\mu$ and/or
$\mu_I\rightarrow-\mu_I$. So it is sufficient to consider only the
region $\mu\ge0$, $\mu_I\ge0$. Taking into account this constraint
and integrating in (\ref{2Omega2}), one obtains the following
expression for the TDP of the system:
\begin{equation}
\begin{split}
\label{2Omega3}
\Omega_{\mu\mu_I}(\sigma,\pi_a)=\frac{\sigma^2+\pi_a^2}{4G}-\int\limits_{-\infty}^{\infty}
\frac{dp_1}{2\pi}\{\varepsilon_++\varepsilon_-+(\mu-\varepsilon_+)\theta(\mu-\varepsilon_+)+
(\mu-\varepsilon_-)\theta(\mu-\varepsilon_-)\},
\end{split}
\end{equation}
where $\theta(x)$ is the Heaviside step function.
To simplify the task, let us note that both the quasiparticle energies
(\ref{epsilon}) and hence the TDP (\ref{2Omega3})  depend effectively
only on two combinations $(\pi_1^2+\pi_2^2)$ and $(\pi_3^2+\sigma^2)$
of the bosonic fields, which are invariants with respect to the
$U_{I_3}(1)\times U_{AI_3}(1)$ group, as is easily seen from
(\ref{inv}). In this case, without loss of generality, one can put
$\pi_2=\pi_3=0$, and study the thermodynamic potential (\ref{2Omega3})
TDP as a function of only two variables, $M\equiv\sigma$ and
$\Delta\equiv\pi_1$. Then the TDP  has the following form:
\begin{equation}
\begin{split}
\label{2Omega4}
\Omega_{\mu\mu_I}(M,\Delta)=\frac{M^2+\Delta^2}{4G}-\int\limits_{-\infty}^{\infty}\frac{dp_1}{2\pi}\{E^+_\Delta+E_\Delta^-+(\mu-E^+_\Delta)\theta(\mu-E^+_\Delta)+(\mu-E^-_\Delta)\theta(\mu-E^-_\Delta)\},
\end{split}
\end{equation}
where
\begin{equation}
\begin{split}
\label{E}
E_\Delta^\pm=\sqrt{\left(\sqrt{p_1^2+M^2}\pm\frac{\mu_I}2\right)^2+\Delta^2}\equiv\sqrt{(E\pm\nu)^2+\Delta^2}  ;\quad\left(E=\sqrt{p_1^2+M^2}, \nu=\frac{\mu_I}{2}\right).
\end{split}
\end{equation}

Since we are going to study the phase diagram of the initial NJL$_2$
model, the system of gap equations is needed:
\begin{equation}
\begin{split}
\label{Gapeq} \frac{\partial\Omega_{\mu\mu_I}(M,\Delta)}{\partial
M}=0;\quad \frac{\partial\Omega_{\mu\mu_I}(M,\Delta)}{\partial
\Delta}=0.
\end{split}
\end{equation}
The coordinates $M$ and $\Delta$ of the global minimum point (GMP)
of the TDP (\ref{2Omega4}) provide two order parameters (gaps),
which are proportional to the ground state expectation values $<\bar
q q>$ and $<\bar q i\gamma^5\tau_1q>$, respectively. In this case
the gap $M$ is just a dynamical quark mass. Obviously, the pair of
gaps $M$ and $\Delta$ is a solution of the system (\ref{Gapeq}). So,
the GMP of the form $(M=0,\Delta=0)$ corresponds to the phase, in
which the initial $U_{I_{3}}(1)\times U_{AI_{3}}(1)$ symmetry
remains intact. If $M\ne 0,\Delta=0$ in the GMP, then the
$U_{I_{3}}(1)$ symmetric phase is realized in the model. Finally,
the GMP of the form $(M=0,\Delta\ne 0)$ corresponds to the
$U_{AI_{3}}(1)$ symmetric phase, where charged pions are condensed
and isospin symmetry, $U_{I_{3}}(1)$, is broken spontaneously. (Note
that due to the zero value of the bare quark mass in the Lagrangian
(1), the GMP of the form $(M\ne 0,\Delta\ne 0)$ does not appear for
the TDP (\ref{2Omega4}).) On the basis of the gaps $M$ and $\Delta$
the following quantities can be introduced,
\begin{equation}
\begin{split}
\label{displaw}
p_{0u}=E_\Delta^--\mu,\quad p_{0d}=E_\Delta^+-\mu,
\quad p_{0\overline{u}}=-(E_\Delta^++\mu) \quad
p_{0\overline{d}}=-(E_\Delta^-+\mu),
\end{split}
\end{equation}
which are the energies of $u$-, $d$-, $\bar u$-, $\bar
d$-quasiparticles (quarks) or one-fermion excitations of the
corresponding phase. The quantities $p_{0u}, p_{0d}$ from
(\ref{displaw}) are the energies necessary for creation of $u$ and
$d$ quarks with momentum $p_1$, whereas $p_{0\overline{u}},
p_{0\overline{d}}$ are the energies necessary for annihilation of
$\bar u$ and $\bar d$ antiquarks.

It is clear that the TDP (\ref{2Omega4}) is an ultraviolet divergent
quantity, so one should renormalize it, using a special dependence
of the bare quantities, such as the bare coupling constant $G\equiv
G(\Lambda)$, on the cutoff parameter $\Lambda$ ($\Lambda$ restricts
the integration region in the divergent integrals, $|p_1|<\Lambda$).
The detailed discussion of this procedure was performed, for
example, in papers \cite{ekzt,massive}. The main stages of it are
the following. First, we transform the expression for the TDP
(\ref{2Omega4}),
\begin{equation}
\begin{split}
\label{2Omega5}
\Omega_{\mu\mu_I}(M,\Delta)=V_0(M,\Delta)-\int\limits_{-\infty}^{\infty}\frac{dp_1}{2\pi}\{E^+_\Delta+
E^-_\Delta-2\sqrt{p_1^2+M^2+\Delta^2}+(\mu-E^+_\Delta)\theta(\mu-E^+_\Delta)+
(\mu-E^-_\Delta)\theta(\mu-E^-_\Delta)\}.
\end{split}
\end{equation}
The integral in the above expression is convergent and $V_0(M,\Delta)$ looks
like
\begin{equation}
\begin{split}
\label{k1}
V_0(M,\Delta)\equiv\Omega_{\mu\mu_I}(M,\Delta)\Big|_{\mu=\mu_I=0}=\frac{M^2+\Delta^2}{4G}-
2\int\limits_{-\infty}^{\infty}\frac{dp_1}{2\pi}\sqrt{p_1^2+M^2+\Delta^2}.
\end{split}
\end{equation}
Second, it is evident that the effective potential (\ref{k1})
\footnote{Recall, the thermodynamic potential in vacuum, i.e. at
$\mu=\mu_I=0$, is usually called effective potential.} is a
divergent quantity. So, to renormalize it we cut the integration
region in (\ref{k1}), $|p_1|<\Lambda$, and require the bare
coupling constant $G\equiv G(\Lambda)$ to have the following form
\begin{eqnarray}
\frac{1}{2G(\Lambda)}=\frac{2}{\pi}\int_{0}^\Lambda
dp_1\frac{1}{\sqrt{M_0^2+p_1^2}}=\frac{2}{\pi}\ln\left
(\frac{\Lambda+\sqrt{M_0^2+\Lambda^2}}{M_0}\right ). \label{16}
\end{eqnarray}
Then in the limit $\Lambda\to\infty$ one can obtain a finite, i.e.
renormalized, expression for $V_0(M,\Delta)$:
\begin{equation}
\begin{split}
\label{2V01}
V_0(M,\Delta)=\frac{M^2+\Delta^2}{2\pi}\left[\mathrm{ln}\left(\frac{M^2+\Delta^2}{M^2_0}\right)-1\right].
\end{split}
\end{equation}
Since $M_0$ might be considered as a free model parameter, it
follows from (\ref{16}) and (\ref{2V01}) that the renormalization
procedure of the NJL$_2$ model is accompanied by the dimensional
transmutation phenomenon. Indeed, in the initial unrenormalized
expression (\ref{k1}) for $V_0(M,\Delta)$ the dimensionless coupling
constant $G$ is present, whereas after renormalization the effective
potential (\ref{2V01}) is characterized by a dimensional free model
parameter $M_0$. Moreover, as it is clear from (\ref{2V01}), the
global minimum point of the affective potential $V_0(M,\Delta)$ lies
just at the point $M=M_0$. So in vacuum, the chiral $SU_L(2)\times
SU_R(2)$ symmetry of the NJL$_2$ model (1) is always spontaneously
broken and the quantity $M_0$ might be treated as a dynamical quark
mass (in vacuum).

Now, taking into account the expression (\ref{2V01}) for
$V_0(M,\Delta)$, we see that the TDP (\ref{2Omega5}) is a finite
renormalization invariant quantity which describes the properties of
dense and isotopically asymmetric quark matter in an infinite
volume.

For the further analysis we need also the expressions for the quark
number density $n_q$ and isospin density $n_I$ in the phase with
gaps $M$ and $\Delta$ which follow directly from (\ref{2Omega5}),
\begin{equation}
\begin{split}
\label{density22}
&n_q\equiv-\frac{\partial\Omega_{\mu\mu_I}}{\partial\mu}=\int\limits_{-\infty}^{\infty}\frac{dp_1}{2\pi}\Big\{\theta(\mu-E_\Delta^+)+\theta(\mu-E_\Delta^-)\Big\},\\
&n_I\equiv-\frac{\partial\Omega_{\mu\mu_I}}{2\partial\nu}=\frac12\int\limits_{-\infty}^{\infty}\frac{dp_1}{2\pi}\Big\{\frac{E+\nu}{E^+_\Delta}\theta(E^+_\Delta-\mu)-\frac{E-\nu}{E^-_\Delta}\theta(E^-_\Delta-\mu)\Big\},
\end{split}
\end{equation}

\subsection{Compactification of the spatial coordinate}

Since physical effects generally take place in a restricted space
region, in the present paper we are going to study the influence of
a finite size of the system (at zero temperature) on the
thermodynamical properties of dense and isotopically asymmetric
quark matter. To simulate a real situation, we put our
(1+1)-dimensional system with Lagrangian (\ref{2LagrGN1}) into a
restricted space region of the form $0\le x\le L$ (here $x$ is the
space coordinate). Then a corresponding modification of the TDP
(\ref{2Omega5}) is needed.

It is well-known that such a constraint on any physical system is
equivalent to its investigation in a spacetime with nontrivial
topology in which space coordinates are compactified. In our case it
means that we can consider the model (1) in spacetime with the
topology $R^1\times S^1$ and with quantum fields satisfying some
boundary conditions of the form
\begin{equation}
\begin{split}
q(t,x+L)=e^{i\pi\alpha}q(t,x),
\end{split}\label{constraint}
\end{equation}
where $0\le\alpha\le2, L$ is the length of the circumference $S^1$,
and the variable $x$ means the path along it. Below, we shall use
only two values of the parameter $\alpha: \alpha=0$ for a periodic
boundary condition and $\alpha=1$ for the antiperiodic one.

As a consequence, to obtain the TDP $\Omega_{L\mu\mu_I}(M,\Delta)$
of the initial system placed in the restricted domain $0\le x\le L$
and at nonzero chemical potentials $\mu\ne 0,\,\,\mu_I\ne 0$, one must simply
replace the integration in (\ref{2Omega5}) and (\ref{k1}) by an
infinite series, according to the rule:
\begin{equation}
\begin{split}
\int\limits_{-\infty}^{\infty}\frac{dp_1}{2\pi}f(p_1)\rightarrow\frac1L\sum\limits_{n=-\infty}^{\infty}
f(p_{1n}); \quad p_{1n}=\frac{\pi}{L}(2n+\alpha), \quad n=0,\pm1,\pm2,\dots
\end{split}
\end{equation}
As a result, we obtain for the corresponding TDP
$\Omega_{L\mu\nu}(M,\Delta)$ in the spacetime of the form $R^1\times
S^1$ the following expression
\begin{equation}
\begin{split}
\label{2Omega6}
\Omega_{L\mu\mu_I}(M,\Delta)=V_L(\rho)&-\frac1L\sum\limits_{n=-\infty}^{\infty}\Big\{E^+_{L\Delta
n}+E^-_{L\Delta n}
-2\sqrt{\rho^2+\frac{\pi^2}{L^2}(2n+\alpha)^2}\\
&+(\mu-E^+_{L\Delta n})\theta(\mu-E^+_{L\Delta n})+(\mu-E^-_{L\Delta
n})\theta(\mu-E^-_{L\Delta n})\Big\},
\end{split}
\end{equation}
where $\rho=\sqrt{M^2+\Delta^2}$,
\begin{equation}
\begin{split}
\label{EL} E_{L\Delta n}^\pm=
\sqrt{\left(\sqrt{M^2+\frac{\pi^2}{L^2}(2n+\alpha)^2}\pm\nu\right)^2+\Delta^2}
\end{split}
\end{equation}
and
\begin{eqnarray}
V_L(\rho)&-&V_L(0)=-\frac{\rho^2} {\pi} \ln\left
(\frac{M_0L}{4\pi}\right )-
\frac{\rho^2\gamma}{\pi}-\frac{2}{L^2}\sqrt{\rho^2L^2+\pi^2
\alpha^2}+\frac{2\pi\alpha}{L^2}\nonumber\\
&-&\frac{2}{L^2}\sum_{n=1}^{\infty}\left
[\sqrt{\pi^2(2n+\alpha)^2+L^2\rho^2}
+\sqrt{\pi^2(2n-\alpha)^2+L^2\rho^2}-4n\pi-\frac{\rho^2L^2}{2n\pi
}\right ]. \label{A9}
\end{eqnarray}
Here $\gamma =0.577...$ is the Euler constant. The expression
(\ref{A9}) is the generalization of the vacuum effective potential
(\ref{2V01}) in the spacetime with trivial $R^1\times R^1$-topology
to the case of spacetime with nontrivial topology of the form
$R^1\times S^1$ (the detailed derivation of (\ref{A9}) is presented
in Appendix of \cite{ekzt}). For further discussion, we need also
the expression for the quark number density $n_{qL}$ in the
$R^1\times S^1$ spacetime, which can be easily obtained from the TDP
(\ref{2Omega6}),
\begin{equation}
\begin{split}
\label{densityL}
&n_{qL}\equiv-\frac{\partial\Omega_{L\mu\mu_I}}{\partial\mu}=
\frac1L\sum\limits_{n=-\infty}^{\infty}\Big\{\theta(\mu-E^+_{L\Delta n})+
\theta(\mu-E^-_{L\Delta n})\Big\}.
\end{split}
\end{equation}
Moreover, in what
follows it will be convenient to use the dimensionless quantities:
\begin{equation}
\begin{split}
\lambda=\frac{\pi}{LM_0},\quad
\tilde{\mu}=\frac{\mu}{M_0},\quad\tilde\nu=
\frac{\nu}{M_0}\equiv\frac{\mu_I}{2M_0}, \quad m=\frac{M}{M_0},\quad
\delta= \frac{\Delta}{M_0}. \label{not}
\end{split}
\end{equation}

\section{Phase structure of the model}

In the present section we shall study the phase structure of the
NJL$_2$ model (1) depending on three external parameters,
$\tilde\mu,\tilde\nu$ and $\lambda$ (\ref{not}) in the case  $T=0$.
In general, this first means that for each fixed values of
$\tilde\mu,\tilde\nu$ and $\lambda$ we should study the global
minimum point of the TDP $\Omega_{\lambda\mu\mu_I}(M,\Delta)$ as a
function of $M,\Delta$ (or dimensionless variables $m,\delta$
(\ref{not})) and then indicate the symmetry group of this point.
Secondly, we must divide the three-dimensional space $(\tilde\mu\ge
0,\tilde\nu\ge 0,\lambda\ge 0)$ into regions (phases), where each of
them is composed by the points corresponding to the same symmetry
group of the TDP global minimum point. Since this is a rather
difficult problem, one can use an equivalent way. Instead of
studying the above mentioned three-dimensional phase diagram, we
will slice it into the planes $\lambda = const$ and then perform the
above procedure only in the planes labeled by some values of
$\lambda$. (Of course, the more values of $\lambda$ are taken into
consideration, the more exact is the obtained picture about the
phase structure of the model.)

\subsection{Particular case: $R^1\times R^1$ spacetime $(\lambda =0)$}
\label{IIIA}

Let us begin our study of the $(\tilde\nu,\tilde\mu,\lambda)$-phase
diagram of the model (1) starting from the plane $\lambda=0$
($L=\infty$) that corresponds to the spacetime of trivial topology
$R^1\times R^1$. In this case the numerical investigations of the
global minimum point (GMP) properties of the TDP (\ref{2Omega5})
bring us to the following phase structure of the model shown in Fig.
1. There, in the corresponding ($\tilde{\nu},\tilde{\mu}$)-plane,
one can see four phase regions denoted by PC as well as by the roman
figures I, II and III, respectively.

In the region denoted by PC, i.e. in the charged pion condensation
phase, each point $(\tilde{\nu},\tilde{\mu})$ corresponds to the GMP
of the form $(m=0,\delta=1)$. Hence, everywhere in this phase the
relation $\tilde\mu< \tilde{\mu}_c<\delta$ is valid,
since  $\tilde{\mu}_c=1/\sqrt2$. As a result, we see from
(\ref{displaw}) that all quasiparticles are
%%%  "gapped" is an unusual jargon, use other word, if possible
gapped in the PC phase, i.e. a finite amount of energy is needed to
create both $u$- and $d$-quarks. Moreover, it is easily seen from
(\ref{density22}) that the quark number density $n_q$ is identically
zero in this phase. In contrast, the isospin density $n_I$ is
nonzero and equal to $\nu/\pi$ at each of the PC phase points.

Then, let us observe the properties of phases denoted by I, II, III.
At each point of these phases the gap $\delta=0$, but the gap $m=0$
appears only in the phases I and III. As a result, in regions I and
III the chirally $U_{I_3}(1)\times U_{AI_3}(1)$-symmetric phase with
massless quarks is arranged. Contrary, in the region II, where the
order parameter $m$ is nonzero, this symmetry is spontaneously
broken down to the isospin $U_{I_3}$ subgroup. In this region the
order parameter $m$ is a smooth function vs $\tilde\mu$ and
$\tilde\nu$ and tends to zero when $\tilde\mu$ and $\tilde\nu$ tend
to infinity ($m\rightarrow0$ when $\tilde{\mu}\rightarrow\infty$ and
$\tilde{\nu}\rightarrow\infty$). We call this phase the normal quark
matter phase, since here quarks dynamically acquire a mass which is
equal to the order parameter $m$ (multiplied by $M_0$), and space
parity is not broken. Note also that the upper and lower boundaries
of the region II tend asymptotically to the line $\tilde\mu=
\tilde\nu$ from the upper- and lower sides, respectively. In the
regions I and III, as it follows from (\ref{density22}), the quark
number density is $n_q=2\mu/\pi$ and the isospin density is $n_I=\nu/\pi$.
Finally, we would like to discuss the difference between I and III
phases. It is evident that in phases I and III the dispersion
relations (\ref{displaw}) for quasiparticles have the following simple
form:
\begin{equation}
\begin{split}
\label{displawm0} p_{0u}=|p_1-\nu|-\mu; \quad p_{0d}=p_1+\nu-\mu.
\end{split}
\end{equation}
Therefore, it is easy to check that in phase I both $u$- and $d$-
quasiparticles are gapless. This means that to create these quarks
costs  no energy, i.e. at each fixed values of $\mu,\nu$ there exist
space momenta $p_1^*$ and $p_1^{**}$ such that $p_{0u}(p_1^*)=0$ and
$p_{0d}(p_1^{**})=0$. In contrast, in phase III only
$u$-quasiparticles are gapless, but $d$-quarks are gapped.
%------------- Fig 1 and Fig 2 ----------
\begin{figure}
\includegraphics[width=0.45\textwidth]{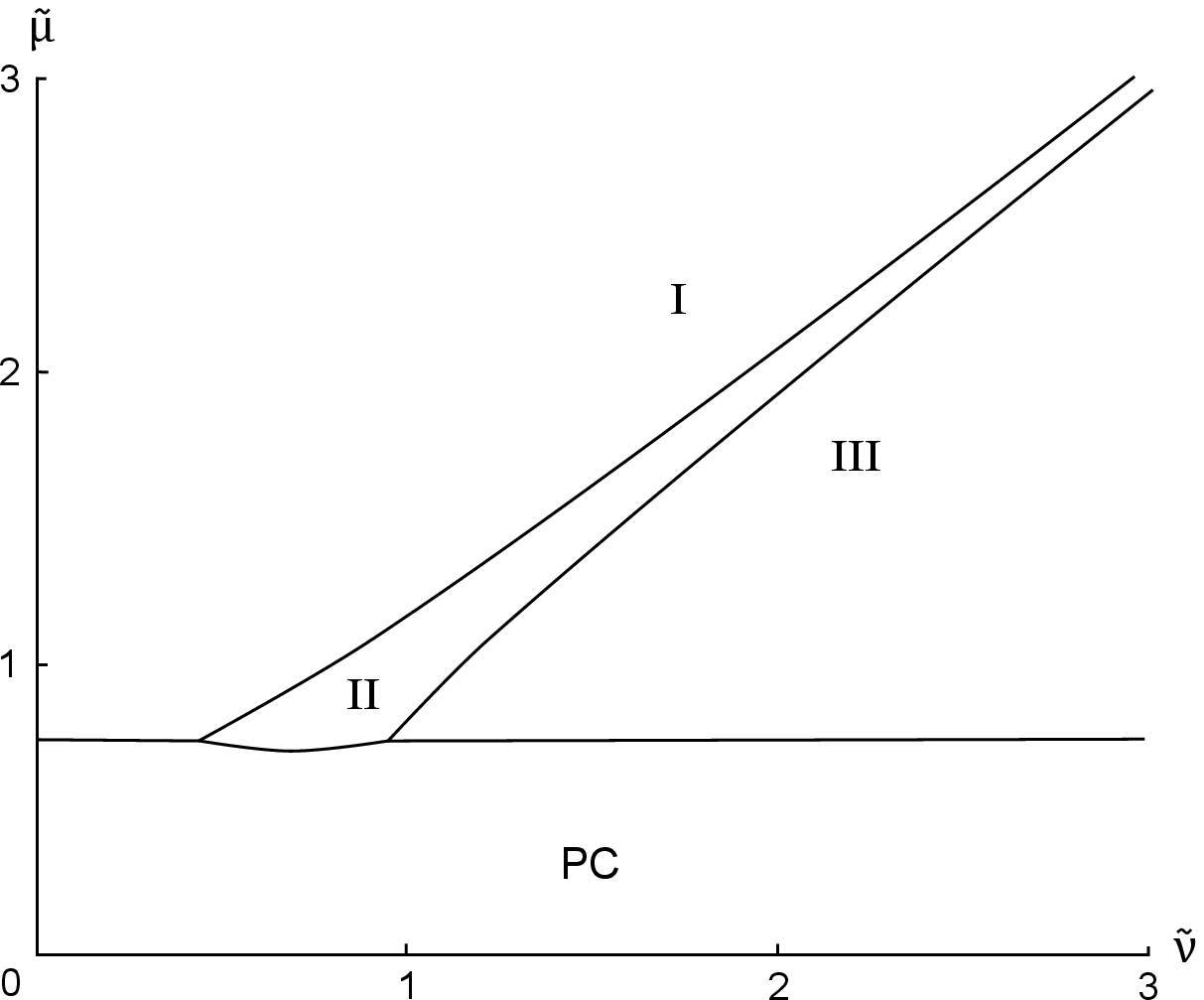}
\hfill
\includegraphics[width=0.45\textwidth]{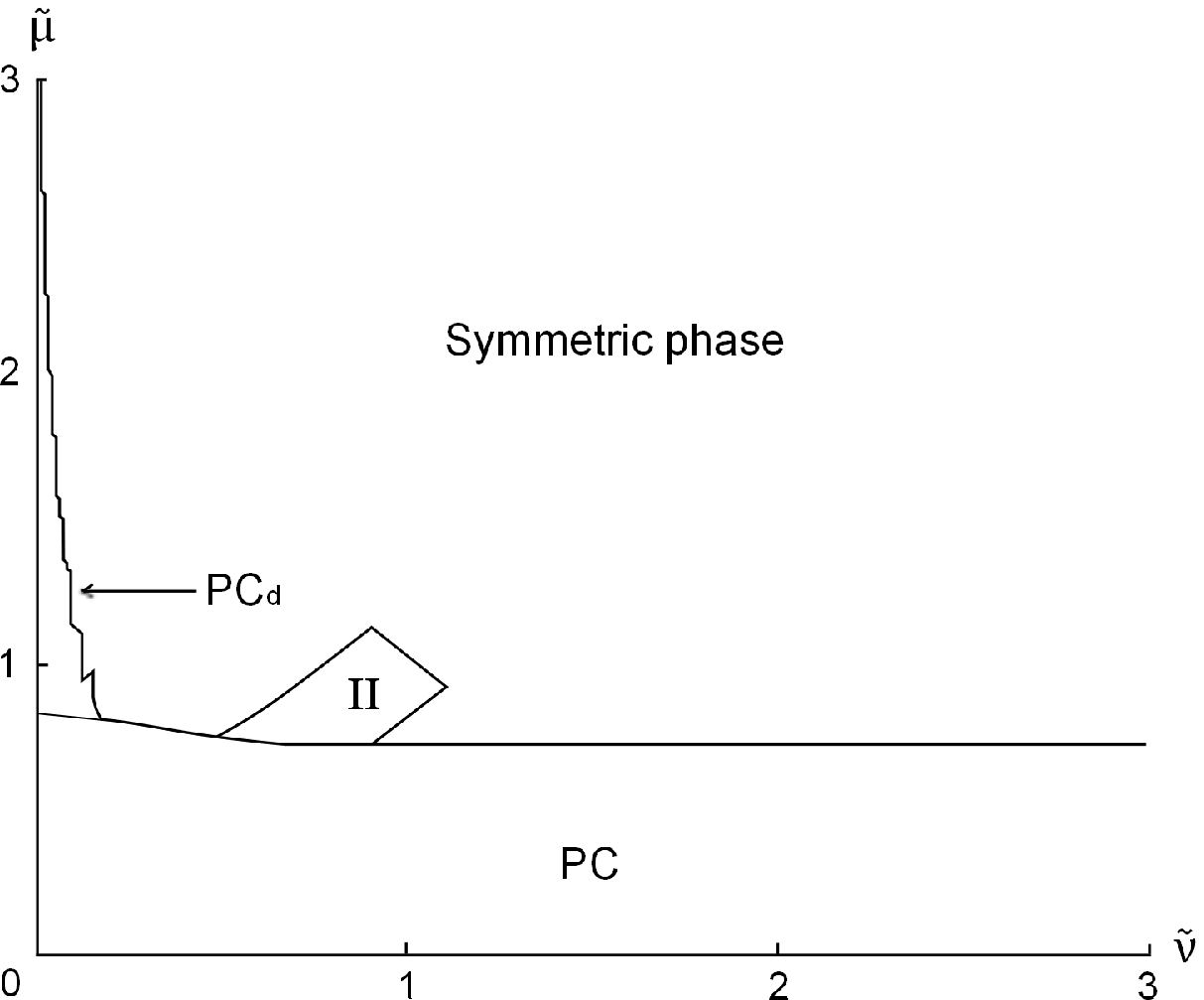}
\\
\parbox[t]{0.45\textwidth}{
 \caption{The $(\tilde\nu,\tilde\mu)$ phase portrait of the model at
$\lambda=0$ and $\tilde\nu>0$. In the phases I and III the initial
$U_{I_3}(1)\times U_{AI_3}(1)$ symmetry is not broken. In the phase
II it is broken down up to $U_{I_3}(1)$, and in the pion
condensation (PC) phase with {\it zero quark number density} it is
spontaneously broken down to $U_{AI_3}(1)$.} } \hfill
\parbox[t]{0.45\textwidth}{
 \caption{{\it The periodic case}: The $(\tilde\nu,\tilde\mu)$ phase portrait at
$\lambda=0.1$. PCd and PC mean the pion condensation phases with
nonzero and zero quark number densities, respectively. II is the
chirally non-symmetrical phase. }}
\end{figure}
%------------- Fig 3 and Fig 4 ----------
\begin{figure}
\includegraphics[width=0.45\textwidth]{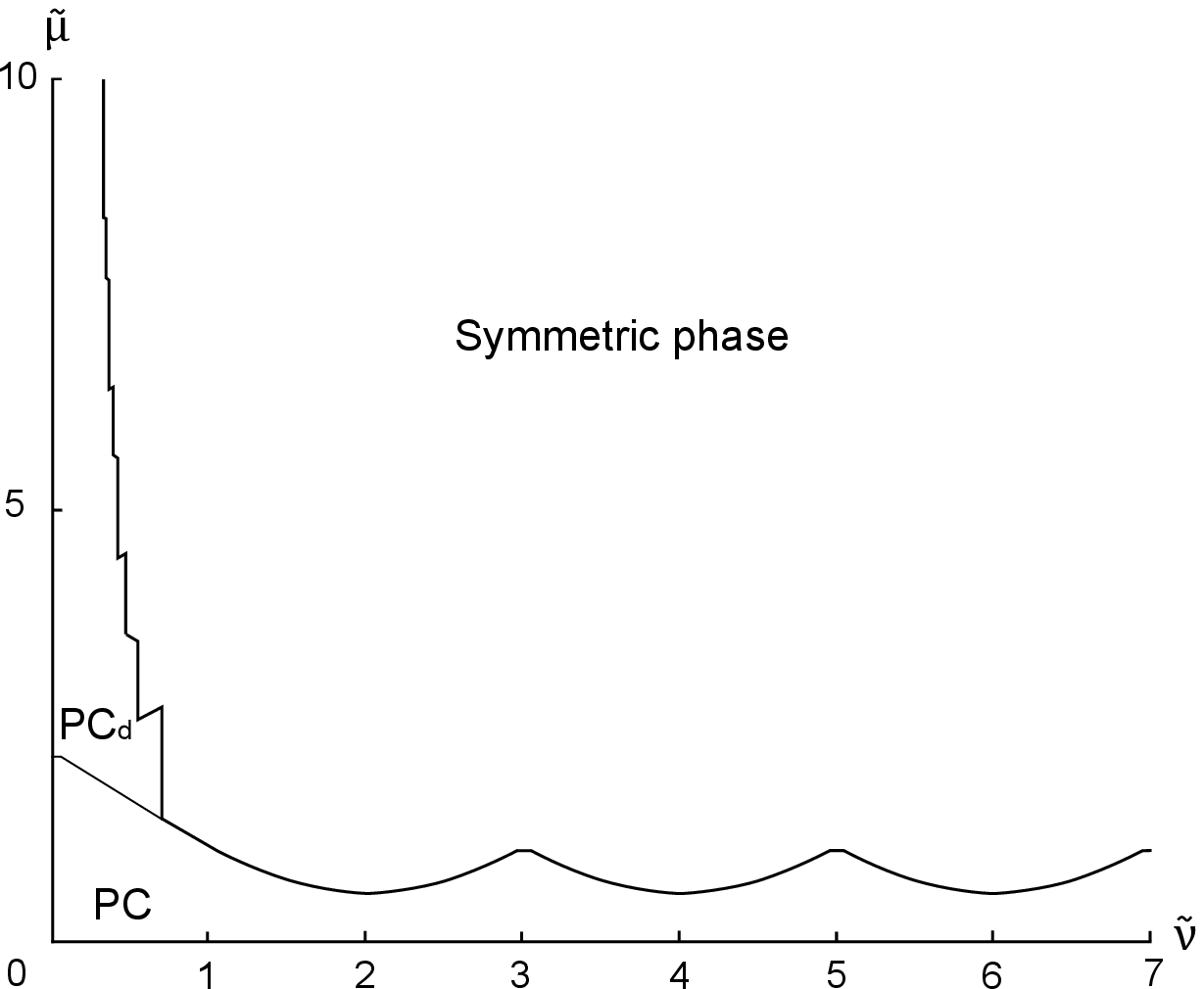}
\hfill
\includegraphics[width=0.45\textwidth]{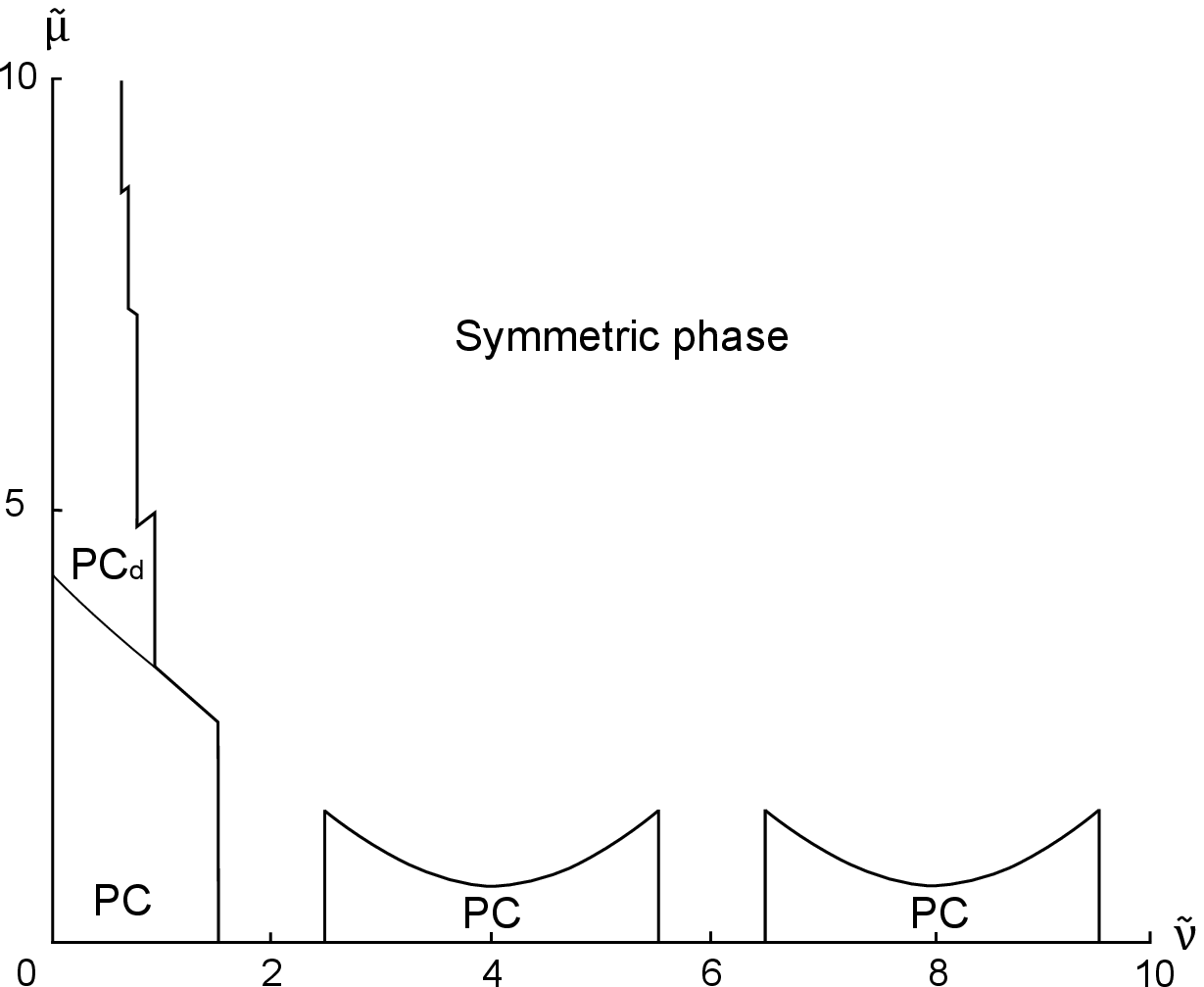}
\\
\parbox[t]{0.45\textwidth}{
 \caption{{\it The periodic case}: The $(\tilde\nu,\tilde\mu)$ phase portrait at
$\lambda=1$. PCd and PC mean the pion condensation phases with
nonzero and zero quark number densities, respectively.} }
\hfill
%%%%%%%%%%%%%%%%%%%%  notation "PC_d" is not so clear/evident %%%%%%%%
%%%%%%%%%%%%%%%555 better could be: $\tilde PC$ or $\bar PC$  ?!
%%%%%%%% If you agree, pleae change it at ALL places !!
%%%%%%%%%%%%%%%%%%%%%%%%%%%%%%%%%%%%%%%%%%%%%%%%%%%%%%%%%%%
\parbox[t]{0.45\textwidth}{
 \caption{{\it The periodic case}: The $(\tilde\nu,\tilde\mu)$ phase portrait at
$\lambda=2.$ PCd and PC mean the pion condensation phases with
nonzero and zero quark number densities, respectively.}}
\end{figure}

\subsection{General case: $R^1\times S^1$ spacetime (periodic boundary conditions)}

In the present section we consider both the phase structure and
relevant properties of the model (1) at some particular nonzero
values of $\lambda$ in the case of periodic boundary conditions,
i.e. at $\alpha=0$ in (\ref{constraint}).  In terms of the
dimensionless quantities (\ref{not}) the TDP (\ref{2Omega6}) has in
the periodic case the following form:
\begin{eqnarray}
\label {per}
&&\frac{\pi}{M_0^2}\Omega_{L\mu\mu_I}(M,\Delta)=(m^2+\delta^2)[\ln(4\lambda)-\gamma]-\lambda\sqrt{(m+\tilde\nu)^2+\delta^2}-\lambda\sqrt{(m-\tilde\nu)^2+\delta^2}\nonumber\\
&&-2\lambda\sum\limits_{n=1}^{\infty}\Big\{ E^+_{n\alpha=0}+E^-_{n\alpha=0}-4n\lambda-\frac{m^2+\delta^2}{2n\lambda}+(\tilde\mu-E_{n\alpha=0}^+)\theta(\tilde\mu-E_{n\alpha=0}^+)+(\tilde\mu-E_{n\alpha=0}^-)\theta(\tilde\mu-E_{n\alpha=0}^-)\Big\},
\end{eqnarray}
where $\gamma =0.577...$ is again the Euler constant and
\begin{equation}
\begin{split}
E_{n\alpha=0}^\pm=\sqrt{(\sqrt{m^2+(2n\lambda)^2}\pm\tilde\nu)^2+\delta^2}.
\end{split}
\end{equation}
We also need  an expression for the quark number density $n_{qL}$ in
the periodic case which follows directly from (\ref{densityL}) at
$\alpha=0$,
\begin{eqnarray}
\label{density0} &&\tilde
n_{qL}\equiv\frac{\pi}{M_0}n_{qL}=\lambda\theta(\tilde\mu-\sqrt{(m+\tilde\nu)^2+\delta^2})+
\lambda\theta(\tilde\mu-\sqrt{(m-\tilde\nu)^2+\delta^2})\nonumber\\&&~~~~~~~~~~~~~~~~~~~~~+
2\lambda\sum\limits_{n=1}^{\infty}\Big\{\theta(\tilde\mu-E^+_{n\alpha=0})+
\theta(\tilde\mu-E^-_{n\alpha=0})\Big\},
\end{eqnarray}
where the zero mode terms are selected in an explicit form. The
results of numerical investigations of the global minimum point
(GMP) of the TDP (\ref{per}) are presented in Figs. 2-4, where the
phase portraits of the model are depicted at $\lambda=0.1,~1,~2$,
respectively. There, in the symmetric phase the global minimum of
the TDP lies at the $U_{I_3}(1)\times U_{AI_3}(1)$ symmetric point
$(m=0,\delta=0)$. The phase II corresponds to the $U_{I_3}(1)$
symmetric GMP of the form $(m\ne 0,\delta=0)$. In this phase quarks
are massive, and the quark number density $n_{qL}$ is nonzero. In
contrast to the case $\lambda=0$ ($L=\infty$), we see that at
$\lambda\ne 0$ the phase II occupies
 a compact region in the phase diagram and completely vanishes
at $\lambda>\lambda_{p}\approx 0.16$.

Moreover, one can see in Figs 2-4 two pion condensation phases, PC
and PCd. They correpond to the   GMP of the TDP (\ref{per}) with
$(m=0,\delta\ne 0)$ and the $U_{AI_3}(1)$ symmetry group.
Clearly, the gap $\delta$ of the pion condensation phases, as well
as the gap $m$ in the above mentioned phase II, depend on the
external parameters, i.e. $\delta\equiv\delta
(\tilde\mu,\tilde\nu,\lambda)$ (see Figs 5, 6). The main difference
between these pion condensation phases is the following. In the PC
phase the quark number density $n_{qL}$ is equal to zero, whereas in
the PCd phase this quantity is nonzero. The boundary between PC and
PCd phases is defined by the zero mode term in (\ref{density0}) at
$m=0$ and $\delta=\delta (\tilde\mu,\tilde\nu,\lambda)$, i.e. it is
the plot of the function $\tilde\mu=\tilde\mu (\tilde\nu,\lambda)$
which is defined implicitly by the equation
\begin{eqnarray}
\label{crit} &&\tilde\mu=\sqrt{\tilde\nu^2+\delta^2
(\tilde\mu,\tilde\nu,\lambda)}.
\end{eqnarray}
%%----------------------------FIGURE 5 and FIGURE 6
\begin{figure}
\includegraphics[width=0.45\textwidth]{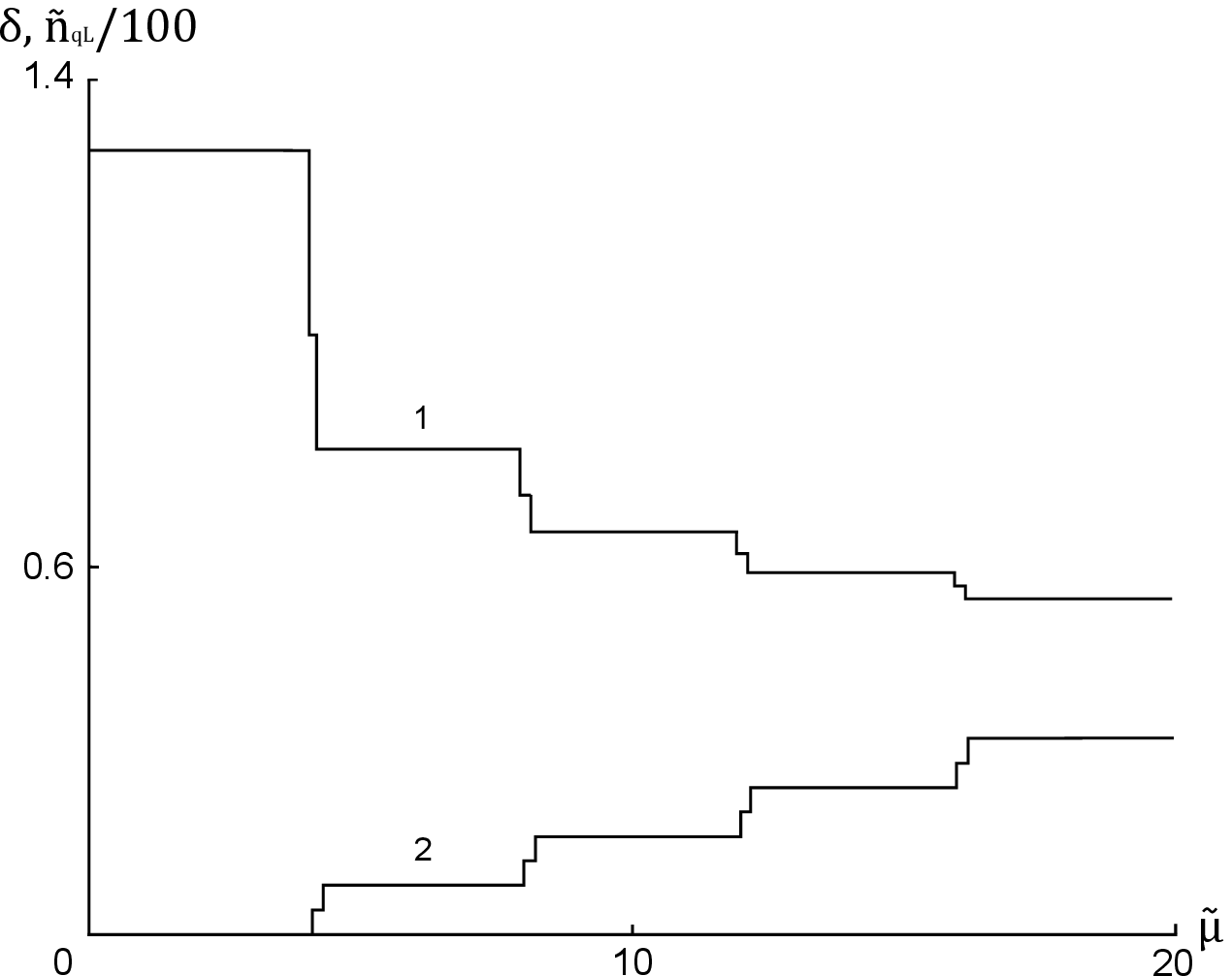}
\hfill
\includegraphics[width=0.45\textwidth]{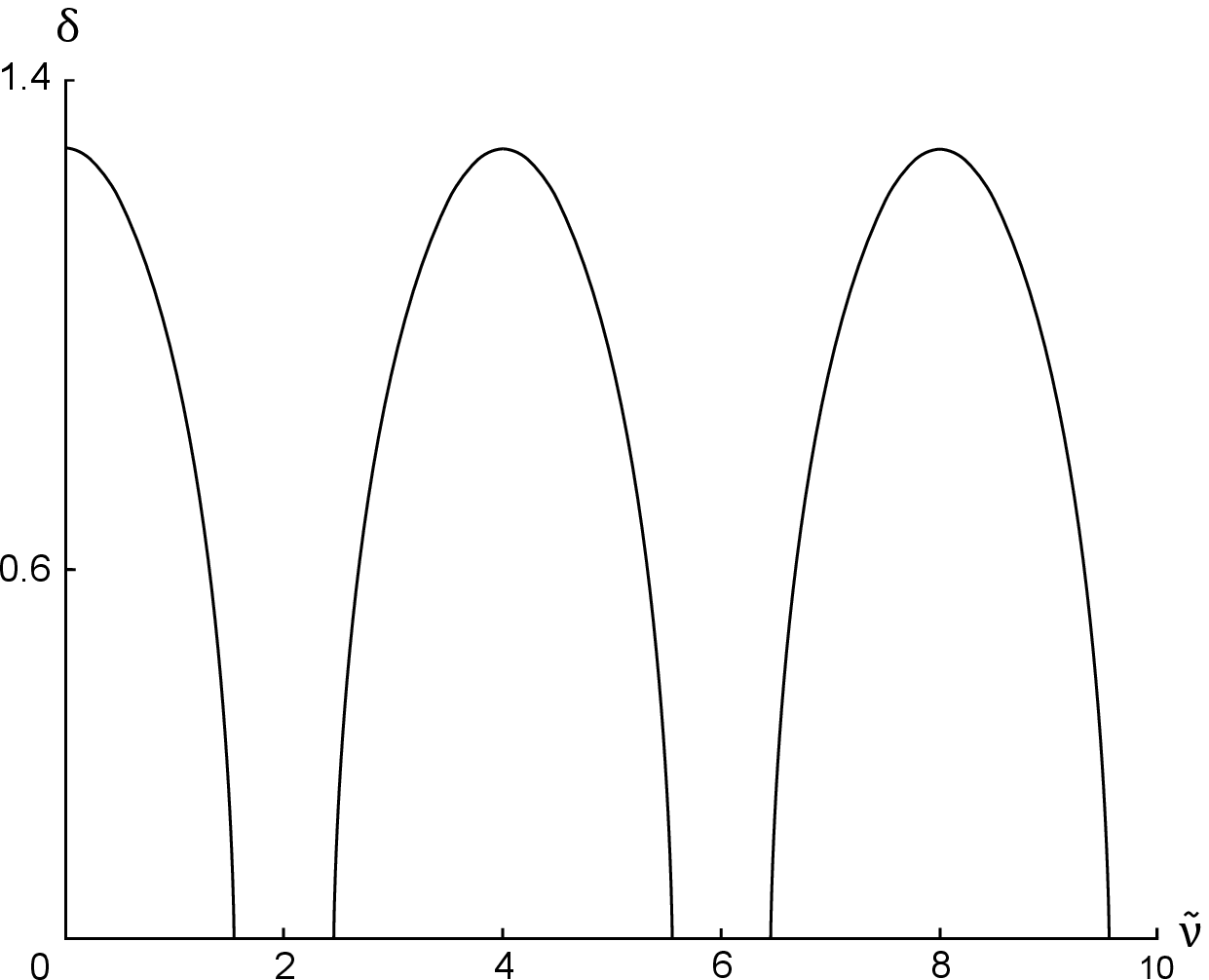}
\\
\parbox[t]{0.45\textwidth}{
 \caption{{\it The periodic case}: The behavior of the gap $\delta$ (curve 1) and quark number density
 $\tilde n_{qL}$ (\ref{density0}) (curve 2) vs $\tilde\mu$ at $\lambda=2$, $\tilde\nu=0.1$.} } \hfill
\parbox[t]{0.45\textwidth}{
 \caption{{\it The periodic case}: The behavior of the gap $\delta$ vs
$\tilde\nu$ at $\lambda=2$ and arbitrary $\tilde\mu\in (0,0.7)$.}}
\end{figure}
Note, we have found the PCd phase at each arbitrary small
$\lambda>0$, whereas at $\lambda=0$, i.e. in the ordinary
(1+1)-dimensional Minkowsky spacetime, it is absent. {\it This means
that a nontrivial spacetime topology promotes the creation of the
charged pion condensed phase with a nonzero quark number density}.
The behavior of both the gap $\delta (\tilde\mu,\tilde\nu,\lambda)$
and quark number density $\tilde n_{qL}$ (\ref{density0}) vs
$\tilde\mu$ in the case of a periodic boundary condition is
illustrated in Fig. 5 at $\lambda=2$ and $\tilde\nu=0.1$.

\subsection{General case: $R^1\times S^1$ spacetime (antiperiodic boundary conditions)}

In the antiperiodic case, i.e. at $\alpha=1$, the TDP
(\ref{2Omega6}) has the following form in terms of variables
(\ref{not}):
\begin{equation}
\begin{split}
\label {anper}
\frac{\pi}{M_0^2}\Omega_{L\mu\mu_I}(M,\Delta)&=(m^2+\delta^2)[\ln(\lambda)-\gamma]-2\lambda\sum\limits_{n=0}^{\infty}\Big\{ E^+_{\alpha=1}+E^-_{\alpha=1}-2(2n+1)\lambda-\frac{m^2+\delta^2}{(2n+1)\lambda}\\
&+(\tilde\mu-E^+_{\alpha=1})\theta(\tilde\mu-E^+_{\alpha=1})+(\tilde\mu-E^-_{\alpha=1})\theta(\tilde\mu-E^-_{\alpha=1})\Big\},
\end{split}
\end{equation}
where
\begin{equation}
\begin{split}
E^\pm_{\alpha=1}=\sqrt{(\sqrt{m^2+(2n+1)^2\lambda^2}\pm\tilde\nu)^2+\delta^2}.
\end{split}
\end{equation}
The investigation of the TDP (\ref{anper}) leads to the following
conclusions about the phase structure of the model (1) (see Figs
7--9): First, as in the case with periodic boundary conditions for
quark fields, at $\lambda\ne\infty$ the chirally non-symmetrical
phase II in the antiperiodic case occupies a compact region in the
$(\tilde\nu,\tilde\mu)$--plane. In addition, this phase is realized
only at $\lambda<\lambda_a\approx 0.66$. Secondly, in the
antiperiodic case the phase portrait of the model contains only the
pion condensed phase with zero quark number density at arbitrary
$\lambda$--values, i.e. a PC phase with nonzero quark number
densities is absent. The behavior of the gap $\delta$ vs $\tilde\nu$
at $\lambda=2$ and arbitrary fixed value of $\tilde\mu\in (0,1)$ is
depicted in Fig. 10.

Finally, note that the above results, obtained both in the periodic
and antiperiodic cases, refer strictly to the case $\nu>0$. A
detailed study of GN-type models in spacetime with nontrivial
topology in the case of $\mu\ne 0$, $\nu=0$ was made in \cite{kim}.
\begin{figure}
\includegraphics[width=0.45\textwidth]{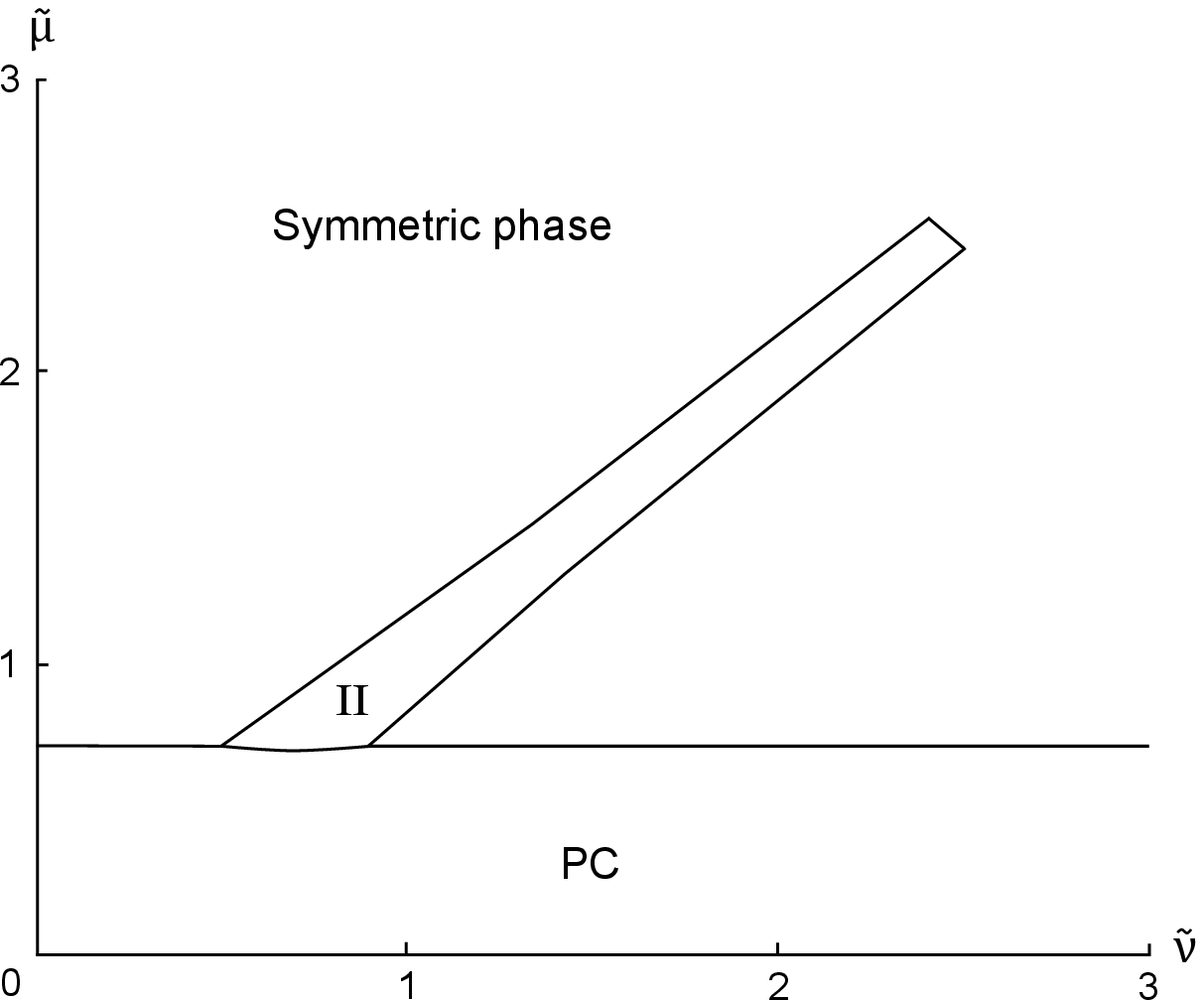}
\hfill
\includegraphics[width=0.45\textwidth]{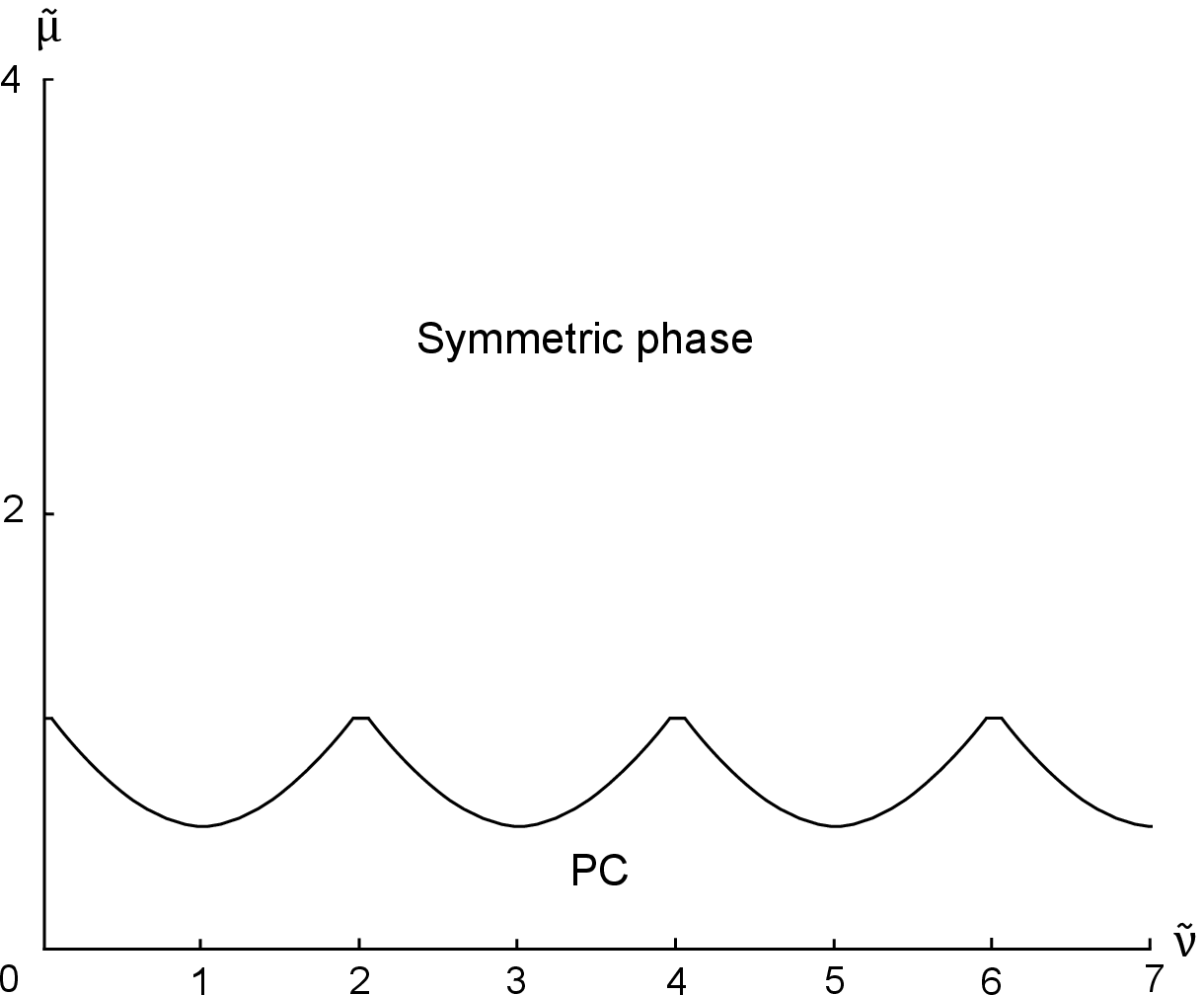}
\\
\parbox[t]{0.45\textwidth}{
 \caption{{\it The antiperiodic case}: The phase portrait at $\lambda=0.1$. PC
again means the pion condensation phase
 with zero quark number densities. II is the chirally nonsymmetric phase.} } \hfill
\parbox[t]{0.45\textwidth}{
 \caption{\textit{The antiperiodic case}: The phase portrait at $\lambda=1$. PC means the pion condensation phase with zero quark number densities.}}
\end{figure}
\begin{figure}
\includegraphics[width=0.45\textwidth]{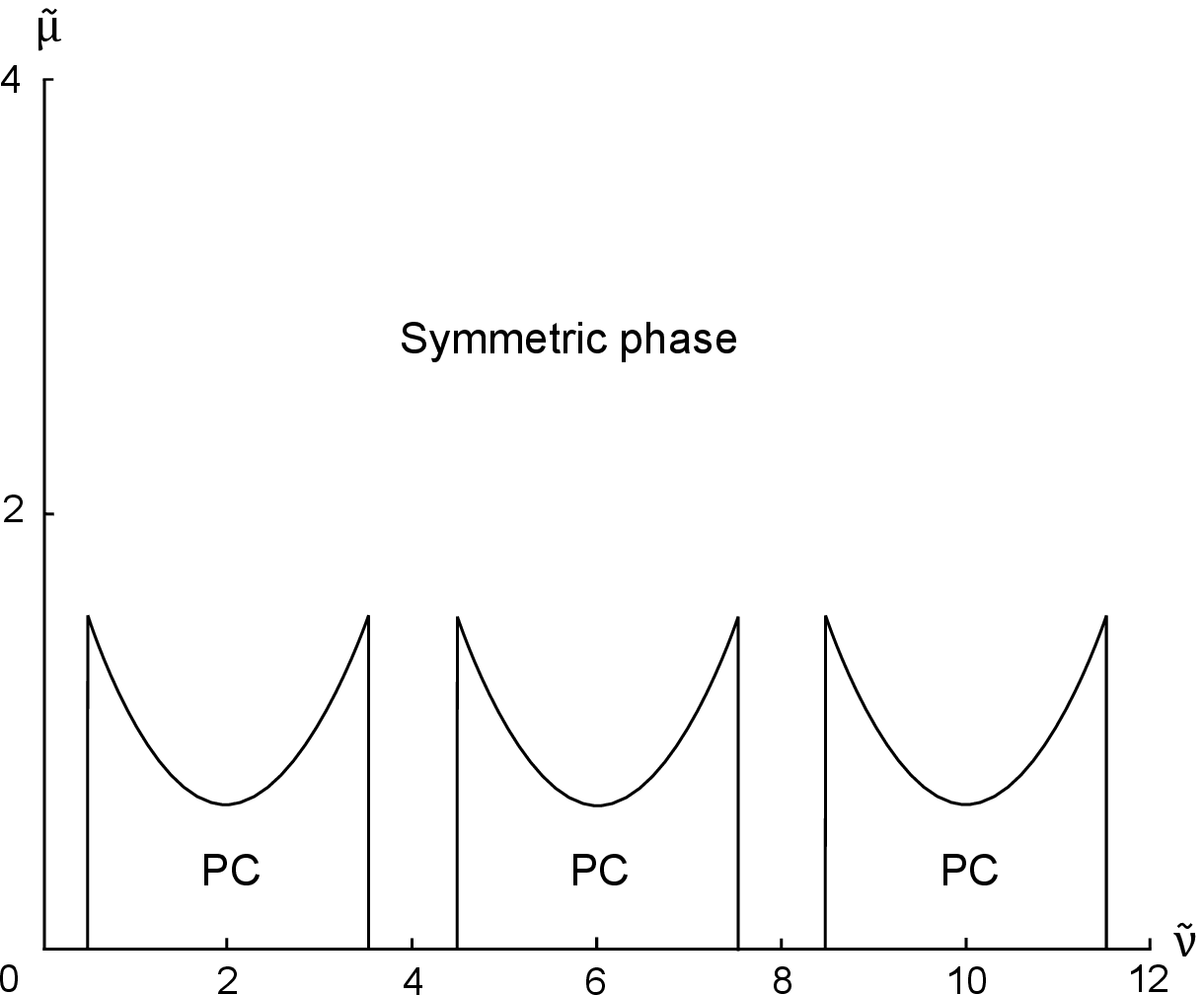}
\hfill
\includegraphics[width=0.45\textwidth]{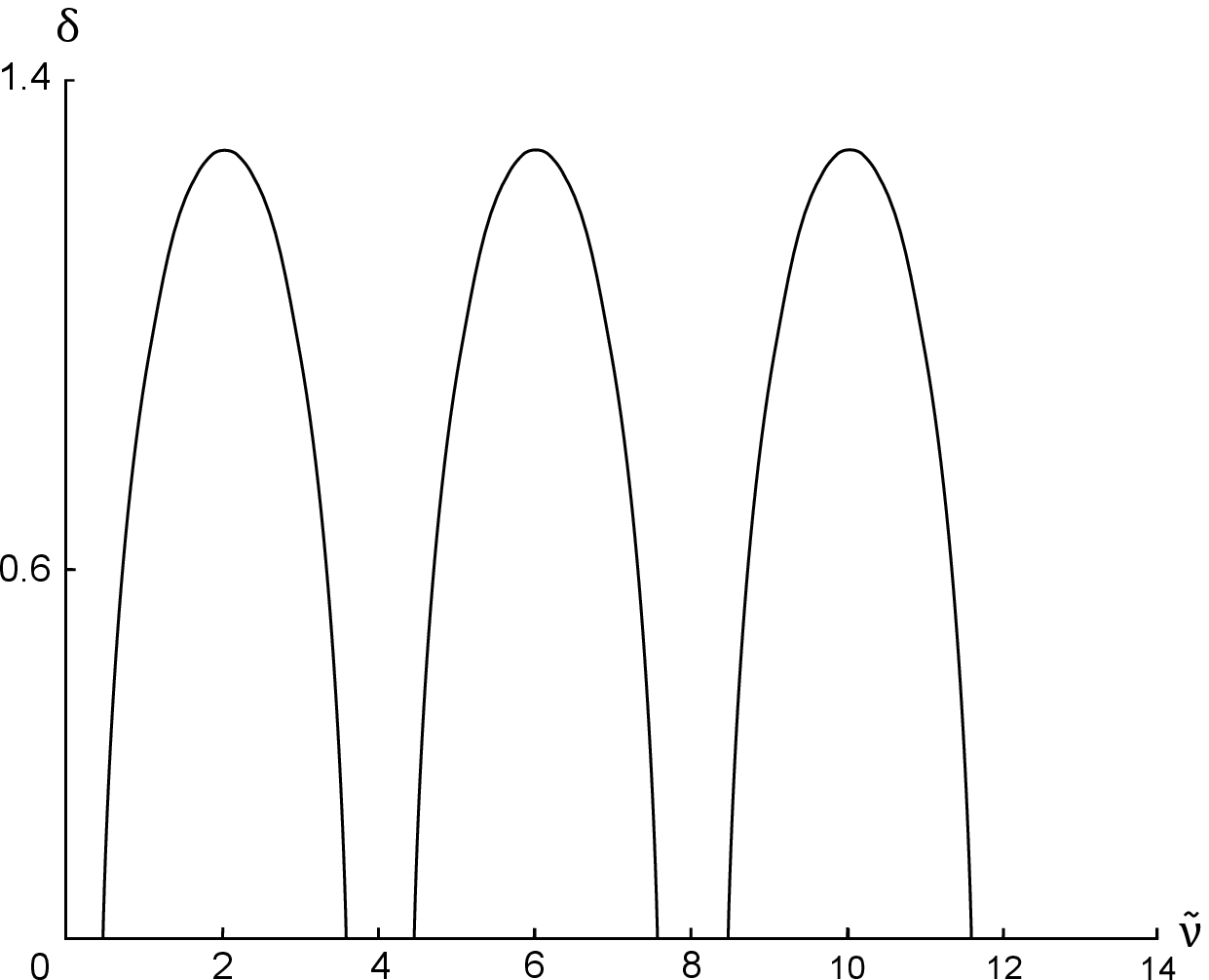}
\\
\parbox[t]{0.45\textwidth}{
 \caption{{\it The antiperiodic case}:  The phase portrait at $\lambda=2$. PC means the pion condensation phase
 with zero quark number densities.} } \hfill
\parbox[t]{0.45\textwidth}{
 \caption{\textit{The antiperiodic case}: The behavior of the gap $\delta$ vs
$\tilde\nu$ at $\lambda=2$ and arbitrary $\tilde\mu\in (0,0.7)$.}}
\end{figure}

\section{Summary and discussion}

In this paper we have studied the so-called charged pion
condensation (PC) phenomenon which might be observed in dense
baryonic matter. Since we were going to pay special attention to the
role of finite volume and to concentrate our attention on the
principle properties of this effect, we have restricted ourselves to
the (1+1)-dimensional NJL$_2$ model (1) with finite values of quark
number $\mu$ and isospin $\mu_I$ chemical potentials at zero
temperature. Moreover, our consideration was performed in the
leading order of the large $N_c$-expansion.

Recall that the charged PC phenomenon was studied recently in the
framework of some QCD-like effective theories such as NJL-type
models or chiral effective theories in usual (3+1)-dimensional
Minkowski spacetime \cite{son,ek,ak,andersen}. However, the
existence of the PC phase with {\it nonzero baryon (quark number)
density}, which we have called the PCd phase, was %not
%%clearly
%detailed enough
%%%%%%%%%%
%described there.
predicted there without any confidence.
% in terms of these effective theories.
Indeed, for some values of model parameters (the coupling constant $G$ and
cutoff parameter $\Lambda$, etc.) the PCd phase
is allowed by the NJL-type models. However, for other physically
interesting values of $G$ and $\Lambda$ the PCd phase is forbidden
in the framework of these models \cite{ek}. Moreover, if the
electric charge neutrality constraint is imposed, the pion
condensation phenomenon depends strongly on the bare (current) quark
mass values. In particular, it turns out that the PCd phase is
forbidden in the framework of the NJL-type models if bare quark
masses reach the physically acceptable values of $5\div 10$ MeV (see
\cite{andersen}). In addition, recent investigations of the charged
pion condensation phenomenon in terms of the (1+1)-dimensional
massive/massless NJL$_2$ model in $R^1\times R^1$ spacetime shows that
the PCd phase is also absent there \cite{massive} (see also Sec.
\ref{IIIA} of the present paper).

In this paper we have placed the  NJL$_2$ system (1)
into a restricted spatial region or, equivalently, in the spacetime
with nontrivial topology of the form $R^1\times S^1$, where the
space coordinate is compactified into a circumference of a finite
length $L$. It turns out that at $\lambda=0$ ($\lambda\sim 1/L$) the
$(\mu_I,\mu)$-phase diagram of the model contains {\it the zero
quark number density PC phase} as well as two symmetric phases,
where the initial symmetry of the model remains intact, and the
phase II with broken chiral symmetry (see Fig. 1). Then, if
$\lambda>0$, we have observed in the  $(\mu_I,\mu)$-phase diagram a
more physically interesting situation, since at periodic boundary conditions for quark fields
and for rather small values of $\mu_I$ the PCd phase is realized
there (see Figs 2-4). So we may conclude that {\it finite size
promotes the appearence of the PC phase with nonzero baryonic
density} at least in the framework of the NJL$_2$ model (1). This is
one of the main results of our paper and we hope that such an effect
also takes place in the more realistic case of (3+1)-dimensional
spacetime. Taking into account the fact that Cooper pairing of
quarks is also significantly facilitated at $L\ne\infty$ \cite{ek2},
it is reasonable to expect that finite size can affect key
properties of any system in comparison with those obtained from the
$L\to\infty$ limit (see, e.g., the papers
\cite{madsen,fraga,kiriyama}, where this point of view is also
supported).

Among other results of our paper, it is interesting to note that the
chirally broken phase II exists on the $(\mu_I,\mu)$-phase diagrams
only at $\lambda<\lambda_p\approx 0.16$ ($\lambda<\lambda_a\approx
0.66$) in the case of periodic (antiperiodic) boundary conditions. In
contrast, the PC phase (PCd phase in the case of periodic boundary
conditions) is realized at an arbitrary value of $\lambda>0$.

Obviously, it would be also interesting to study the influence of
finite-size effects on the existence of chiral density waves and on
the PC phenomenon in some QCD-like models in (3+1)-dimensional
spacetime.

\section*{Acknowledgments}

Two of the authors (V.Ch.Zh. and T.G.Kh.) are grateful to Professor
M. Muller-Preussker for his kind hospitality during their stay in the
particle theory group at the Institute of Physics of
Humboldt-University, where part of this work has been done, and also
to DAAD for financial support.

\end{document}